\documentclass[letterpaper,10pt]{article} 
\usepackage{opticameet3} %% use version 3 for proper copyright statement

\newcommand\authormark[1]{\textsuperscript{#1}}

\usepackage{amsmath,amssymb}
\usepackage[colorlinks=true,bookmarks=false,citecolor=blue,urlcolor=blue]{hyperref} %pdflatex

\usepackage[nolist]{acronym}
\usepackage{orcidlink}
\usepackage[caption=false,font=normalsize]{subfig} % change font if needed

\acrodef{AE}{autoencoder}
\acrodef{AI}{artificial intelligence}
\acrodef{AI/ML}{artificial intelligence \& machine learning}
\acrodef{ANN}{artificial neural network}
\acrodef{BER}{bit error rate}
\acrodef{CoS}{cosine similarity}
\acrodef{ECL}{external cavity laser}
\acrodef{EDFA}{Erbium-doped fiber amplifier}
\acrodef{GB}{gradient boosting}
\acrodef{GRU}{gated recurrent unit}
\acrodef{LSTM}{long short-term memory}
\acrodef{ML}{machine learning}
\acrodef{MS-SNN}{multi-similarity Siamese neural network}
\acrodef{NB}{naive Bayes}
\acrodef{NLI}{nonlinear interference}
\acrodef{NN}{nearest neighbor}
\acrodef{OPM}{optical performance monitoring}
\acrodef{OSNR}{optical signal-to-noise ratio}
\acrodef{OTDR}{optical time-domain reflectometry}
\acrodef{QoT}{quality of transmission}
\acrodef{RBF}{radial basis function}
\acrodef{RF}{random forest}
\acrodef{SNN}{Siamese neural network}
\acrodef{SSMF}{standard single-mode fiber}
\acrodef{SVM}{support vector machine}

\begin{document}

\title{A Unified Siamese Learning Framework for Zero-Day Anomaly Detection and Classification in Optical Networks}

\author{
Carlos Natalino~\orcidlink{0000-0001-7501-5547},\authormark{1, *}
Flávia P. Monteiro~\orcidlink{0000-0002-8153-472X},\authormark{2}
and Paolo Monti~\orcidlink{0000-0002-5636-9910},\authormark{1}
}

\address{
\authormark{1}Department of Electrical Engineering, Chalmers University of Technology, 412 96 Gothenburg, Sweden\\
\authormark{2}Federal University of Western Pará (UFOPA), 68040-255 Santarém, Pará, Brazil
}

\email{\authormark{*}carlos.natalino@chalmers.se} %% email address is required

%% Do not add a copyright statement. Optica will add it.

\begin{abstract}
A multi-similarity Siamese neural network unifies zero-day anomaly detection and one-shot classification in optical networks, achieving over 99\% accuracy and instant adaptability across lightpaths and unseen anomaly types without any retraining.\\
This is the authors' version of this publication. The final published version is available at \url{https://doi.org/10.1364/OFC.2026.M3A.2}.
\end{abstract}

\section{Introduction}

Optical networks are increasingly complex systems, where anomalies arise from the interplay of heterogeneous devices, configurations, physical impairments, or external factors such as infrastructure degradation \cite{Musumeci2025jocn}.
Reliable operation therefore requires flexible mechanisms for anomaly detection and classification.
However, the massive volume of monitoring data makes manual oversight and threshold-based methods impractical for network-wide deployments, driving operators toward \ac{AI/ML}-based automation~\cite{Chen2022commag,Musumeci2025jocn}.
Yet, models trained for specific lightpaths or device/network configurations often fail to generalize, calling for invariant, network-wide representations capturing the underlying behavior of the optical infrastructure.

A second challenge is the integration of anomaly detection and classification.
Traditional approaches treat detection (i.e., identifying whether an anomaly exists) and classification (i.e., identifying the anomaly type or class) as separate stages.
For example, \cite{Ghosh_CompNet_2025} proposed a \ac{LSTM} \ac{AE} for anomaly detection followed by an \ac{LSTM} classifier for soft-failure identification, while \cite{Abdelli2022jocn} combined an \ac{AE} for detection with an attention-based bidirectional \ac{GRU} for fault diagnosis.
Other works explored domain adaptation and transfer learning to reduce the need for labeled samples \cite{Musumeci2022jocn}.
Although effective, these methods rely on rigid decision boundaries and offer limited transferability, making them unsuitable for dynamic conditions.

An additional difficulty is the zero-day anomaly problem, caused by previously unseen anomaly types not part of the training data.
The anomaly landscape evolves continuously with new technologies, traffic dynamics, or environmental conditions, making static models quickly obsolete.
While anomalies are rare, this scarcity of labeled data hinders robust model development.
Supervised classifiers struggle with novel anomalies without complete retraining, and unsupervised methods can detect but not classify them.
This motivates few-shot or one-shot learning models capable of recognizing and classifying a new anomaly from its first occurrence (i.e., day zero) without retraining.
Recent studies have explored the use of \acp{SNN} and similarity-based learning for such low-sample regimes.
For example, \cite{Gao_2024_oecc} combined an \ac{SNN} with pattern mining to handle imbalanced anomaly data and perform one-shot clustering of new anomaly types, while \cite{Natalino2019ap} applied \acp{SNN} to modulation format classification, achieving strong generalization to unseen classes, but using only a single similarity metric.
However, a single metric may be insufficient to capture the multidimensional dependencies among optical performance parameters, limiting generalization across diverse network conditions.

This paper introduces a \ac{MS-SNN} framework for zero-day anomaly detection and one-shot classification in optical networks.
The framework:
\emph{(i)} generalizes across different lightpaths and network conditions through multi-similarity feature learning,
\emph{(ii)} integrates detection and classification into a unified learning process, and
\emph{(iii)} enables one-shot classification of unseen anomalies from their first observation.
In contrast to \ac{SNN} that rely on a single similarity metric, the proposed \ac{MS-SNN} framework captures a richer representation of relationships among optical performance parameters, thus improving model robustness to unseen anomaly types.
Experiments on an open dataset \cite{Ghosh_CompNet_2025} show over 99\% accuracy in one-shot classification scenarios, demonstrating a major step toward adaptive, network-wide anomaly intelligence in future optical infrastructures.

\section{Zero-day anomaly detection and classification based on \acp{MS-SNN}}

\begin{figure}[htb]
    \centering
    \includegraphics[width=0.95\linewidth, trim={1.7cm 8cm 3cm 5.5cm}, clip]{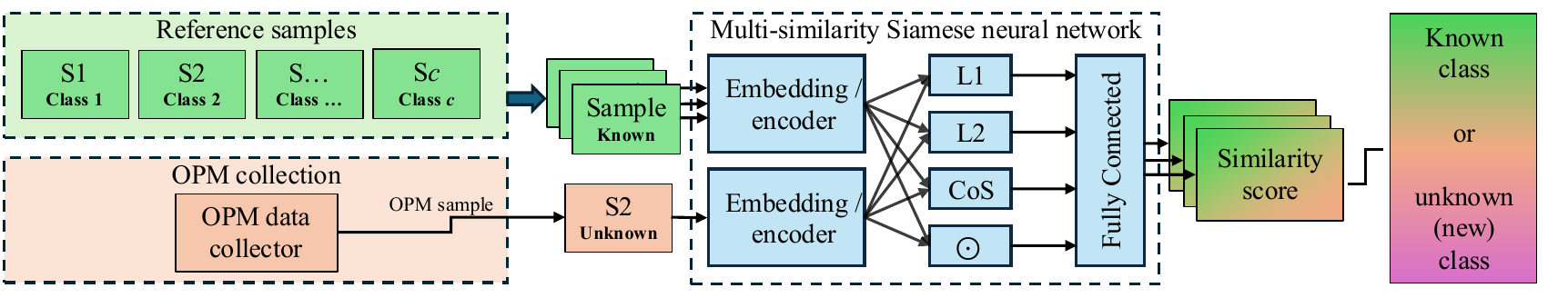}
    % Answer: [trim={left bottom right top},clip]
    \caption{Architecture of the proposed \acf{MS-SNN} with multi-similarity metrics: two samples share the same embedding/encoder segment.}
    \label{fig_architecture}
\end{figure}

Our proposed framework employs a \ac{MS-SNN} designed for generalizable zero-day anomaly detection in optical networks.
Instead of learning a direct classification boundary, the \ac{SNN} learns a similarity function over pairs of samples, i.e., a reference from a known class and a sample of interest.
The \ac{SNN} comprises three main segments: an embedding/encoder, a multi-similarity head, and a similarity computation module.
Each sample is processed by the shared embedding/encoder, which extracts a high-dimensional feature representation.
These representations are compared through a multi-similarity head, which computes complementary distance metrics, followed by a fully-connected layer that outputs a similarity score indicating whether the two samples belong to the same class.

Fig. \ref{fig_architecture} illustrates the proposed framework during operation.
An incoming \ac{OPM} sample is compared against a gallery of reference samples, each representing a known anomaly class.
If the similarity score between the new sample and all known reference samples falls below a calibrated threshold, the system flags it as a new, unknown (or zero-day) anomaly class, which can be labeled afterwards by specialists.
The first sample of this new class is then added to the reference gallery, allowing subsequent samples to be classified without retraining.
This mechanism enables one-shot classification and allows the model to continuously expand its knowledge base.
Although classification complexity scales linearly with the number of classes, parallel inference on modern \ac{AI} accelerators completes all comparisons within milliseconds, making the approach practically feasible for real-time operations.

The embedding/encoder consists of fully-connected layers mapping \ac{OPM} parameters into a high-dimensional feature space.
The multi-similarity head computes four similarity/distance measures (i.e., L1 (Manhattan), L2 (Euclidean), \ac{CoS}, and element-wise product) to capture different relationships between feature vectors.
The similarity module then combines these measures through additional fully connected layers with learnable weights, enhancing the model's ability to capture non-linear correlations among \ac{OPM} parameters.

The network is trained using contrastive learning with a binary cross-entropy loss function, reframing the multi-class problem into a binary decision: whether two samples are similar or not.
During training, sample pairs from the same class are labeled as similar (1), while pairs from different classes are labeled as dissimilar (0).
The objective of the training process is to minimize the binary cross-entropy loss, which effectively teaches the network to generate embeddings that are close together for samples of the same class and far apart for samples from different classes.
This strategy is effective in data-scarce scenarios (i.e., anomaly management in optical networks) because
it combinatorially expands the number of training examples from a small labeled set, allowing the model to learn a robust similarity metric without a large number of instances for each class.
By learning a general notion of similarity rather than fixed decision boundaries, the  \acf{MS-SNN} remains independent of the total number of classes and can generalize to unseen anomalies without modification

Once trained, the same \ac{MS-SNN} supports both anomaly detection and classification.
For zero-day anomaly classification, only one reference sample per class is required.
The class of a new sample is determined by averaging and/or thresholding its similarity scores across all reference samples, enabling efficient one-shot learning and strong generalization across network conditions.

\section{Evaluation of the Proposed Framework}

\begin{figure}[t]
    \centering
    \subfloat[Accuracy]{\includegraphics[width=0.54\linewidth]{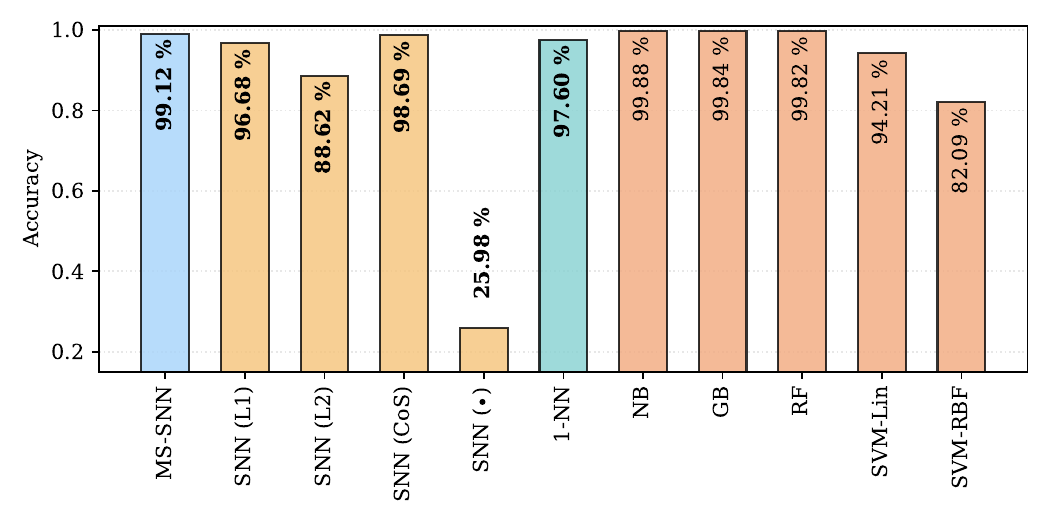}}
    \subfloat[Confusion matrix for MS-SNN]{\includegraphics[width=0.46\linewidth]{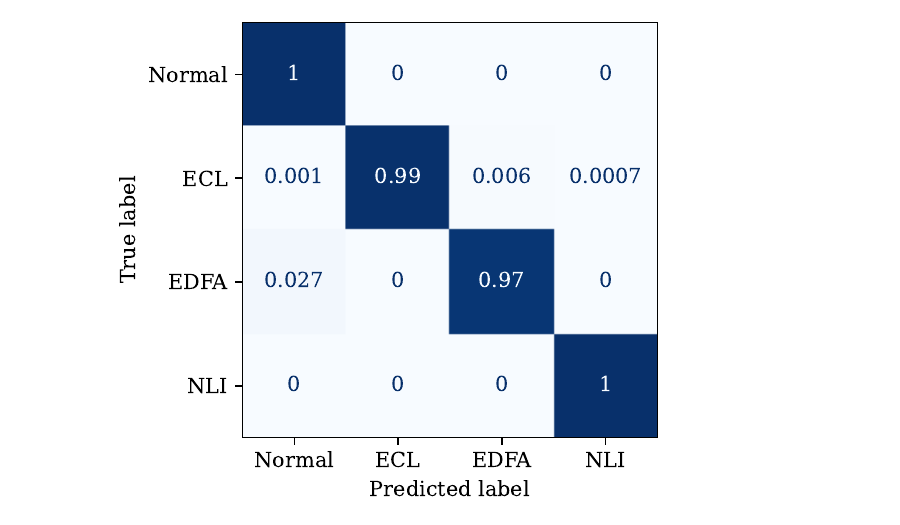}}
    \caption{Performance comparison between single-sample methods (\ac{MS-SNN}, single-similarity SNNs, and 1-NN) and conventional \ac{ML} models (\acf{NB}, \acf{GB}, \acf{RF}, and linear/\acf{RBF} \acp{SVM}) trained on all four anomaly classes.}
    \label{fig_results}
\end{figure}

We use the dataset from \cite{Ghosh_CompNet_2025} to validate the proposed \acf{MS-SNN} framework.
It considers a 28-node, 41 links optical backbone network with inline amplifiers every 80 km and coherent receivers with \ac{OPM} capability, providing \ac{QoT} parameters. 
Each \ac{OPM} sample includes five features: lightpath length, laser current, lightpath received power, \ac{OSNR}, and \ac{BER}.
Samples were collected from 756 lightpaths (one between every source-destination node pair) using the shortest path.
The dataset was generated by simulating three types of soft failures dominant in backbone optical networks:
\ac{ECL} malfunction, failures related to increasing threshold current;
\ac{EDFA} malfunction, failures due to decreasing pump laser power/gain of inline amplifiers; and 
\ac{NLI} soft failures caused by randomly increasing lightpath transmit power.
The \ac{QoT} data comprise 900 time samples (approximately 150 hours) per lightpath collected under normal and faulty conditions.
At any given time, a lightpath can experience at most one failure type.

The \ac{MS-SNN} architecture comprises 5 fully-connected layers with batch normalization, ReLU activation, and 0.3 dropout rate, with 128, 256, 128, 64, and 32 neurons, respectively.
The resulting embeddings are used to compute the four distance metrics discussed earlier: Manhattan, Euclidean, element-wise product, and cosine similarity.
These are concatenated and fed to a fully-connected layer with softmax activation, followed by an output neuron with sigmoid activation.
For training, we used 1,000 samples from all lightpaths and two classes: 0 (normal) and \ac{ECL} malfunction (1).
This deliberately limited training set allows evaluating the model's zero-day detection and classification capability.
Testing used 10,000 samples from each of the four classes to simulate zero-day scenarios, comprising samples from all lightpaths.
The network was trained for 100 epochs using the Adam optimizer (learning rate of 0.001, batch size of 64).

To assess the benefit of multi-similarity learning, we benchmark the proposed \ac{MS-SNN} against four variations of the same \ac{SNN} architecture, each using a single similarity metric.
The \ac{MS-SNN} was also compared with the \ac{NN} algorithm (1-NN), which also relies on one sample per class, and with traditional \ac{ML} models (\ac{NB}, \ac{GB}, \ac{RF}, and linear and \ac{RBF} \ac{SVM}s) that require labeled data for all classes.
These baseline \ac{ML} models were trained with 1,000, and tested over 10,000 samples per class.
Unlike the proposed framework, they may not detect unseen anomalies, and cannot classify unseen anomaly types, being limited to closed-set classification.

Fig. \ref{fig_results} summarizes the results.
The proposed \ac{MS-SNN} achieves 99.12\% accuracy, outperforming single-metric \acp{SNN} and 1-NN (97.6\%) (Fig. \ref{fig_results}(a)).
Among the single-metric variants, the one with \ac{CoS} performs best yet below 99\%, followed by L1 with 96.6\%.
Surprisingly, the \ac{SNN} element-wise product as similarity metric only achieves 25.9\% accuracy, which is very close to the theoretical random classifier performance for four classes (i.e., 25\%).
This confirms that combining multiple similarity measures yields richer, more discriminative representations and stronger generalization to unseen anomaly types.
Traditional ML methods (\ac{NB}, \ac{GB}, and \ac{RF}) achieve slightly higher accuracy (around 99.8\%), but rely on labeled data for all classes, defeating the purpose of zero-day classification. 
This emphasizes the strength of \ac{MS-SNN}: high accuracy with minimal supervision and no retraining.
The confusion matrix in Fig. \ref{fig_results}(b) shows 97\% anomaly detection accuracy, i.e., around 3\% of false negatives, where the failures are misclassified as normal operating conditions.
Notably, the \ac{MS-SNN} perfectly identifies normal operating states, i.e., it produces no false positives, avoiding unnecessary alarm overhead.

\section{Final remarks}

This work proposed an \ac{MS-SNN}-based framework that achieves zero-day anomaly detection and one-shot classification with over 99\% accuracy and no false positives, paving the way for scalable, adaptive anomaly management in future optical networks.
Future works may explore the suitability of the proposed approach over other anomaly detection and classification scenarios, including anomaly identification, and localization.

\section*{Acknowledgments}

This work has been supported by the Horizon Europe ECO-eNET project with funding from the SNS JU under grant agreement No. 101139133.
The implementation is available at \\\url{https://github.com/carlosnatalino/OFC_2026_UnifiedSiameseLearning}.

\bibliographystyle{opticaconf}
\bibliography{references}

\end{document}